\documentclass[pra,twocolumn,a4paper,showpacs,superscriptaddress]{revtex4}
\usepackage{latexsym}
\usepackage{amsmath}
\usepackage{amsfonts}
\usepackage{graphicx}
\usepackage{exscale}
\usepackage{amssymb}
\usepackage{bbold,mathrsfs}
\usepackage{natbib}
\newcommand{\ket}[1]{\ensuremath{|#1\rangle}}
\newcommand{\bra}[1]{\ensuremath{\langle  #1 |}}

\newcommand{\vro}{\varrho}
\newcommand{\be}{\begin{equation}}
\newcommand{\ee}{\end{equation}}
\newcommand{\ba}{\begin{eqnarray}}
\newcommand{\ea}{\end{eqnarray}}

\newcommand{\mc}[1]{\ensuremath{\mathcal{#1}}}
\newcommand{\bc}{\begin{center}}
\newcommand{\ec}{\end{center}}
\newcommand{\bi}{\begin{itemize}}
\newcommand{\ei}{\end{itemize}}
\newcommand{\mean}[1]{\ensuremath{ \langle #1  \rangle}}
\newcommand{\meanb}[1]{\ensuremath{ \big \langle #1 \big\rangle}}
\newcommand{\meanI}[1]{\ensuremath{ \langle #1  \rangle_0}}

\newcommand{\mf}[1]{\boldsymbol{#1}}

\newcommand{\rf}{\vro_{\text{F}}}
\newcommand{\rft}{\tilde{\vro}_{\text{F}}}

\begin{document}

\title{Two-mode single-atom laser as a source of entangled light}

\author{M. \surname{Kiffner}}
\email{martin.kiffner@mpi-hd.mpg.de}
\affiliation{Max-Planck-Institut f\"ur Kernphysik, 
Saupfercheckweg 1, 69117 Heidelberg, Germany}

\author{M.~S. \surname{Zubairy}}
\email{zubairy@physics.tamu.edu}
\affiliation{Max-Planck-Institut f\"ur Kernphysik, 
Saupfercheckweg 1, 69117 Heidelberg, Germany}
\affiliation{\mbox{Institute for Quantum Studies and Department of Physics, Texas A\&M University, 
College Station, Texas 77843, USA}}

\author{J. \surname{Evers}} 
\email{joerg.evers@mpi-hd.mpg.de}
\affiliation{Max-Planck-Institut f\"ur Kernphysik, 
Saupfercheckweg 1, 69117 Heidelberg, Germany}

\author{C.~H. \surname{Keitel}}
\email{keitel@mpi-hd.mpg.de}
\affiliation{Max-Planck-Institut f\"ur Kernphysik, 
Saupfercheckweg 1, 69117 Heidelberg, Germany}


\begin{abstract} 
A two-mode single-atom laser is considered, with the aim of generating entanglement
in macroscopic light. Two transitions in the four-level gain medium atom
independently interact with the two cavity modes, while two other transitions
are driven by control laser fields. Atomic relaxation as well as cavity losses
are taken into account. We show that this system is a source
of macroscopic entangled light over a wide range of control parameters
and initial states of the cavity field.
\end{abstract}

\maketitle

\section{INTRODUCTION}
Quantum entanglement is known to be the key resource in many applications of
quantum information and quantum computing~\cite{nielsen:00}. 
These phenomena range from quantum
teleportation \cite{bennett:93,bouwmeester:97} and quantum cryptography \cite{bennett:84}
to quantum implementation of Shor's algorithm \cite{shor:97} and quantum search
\cite{grover:97}. It is therefore not surprising that there has been a great deal
of interest in the generation and measurement of entanglement in recent years.

Entangled states have been considered traditionally between individual qubits.
However, it has been shown that continuous variable entanglement can offer an
advantage in some situations in quantum information science~\cite{braunstein:05}. 
One reason for this is that continuous variable entanglement often can be
prepared unconditionally, whereas the preparation of discrete entanglement
usually relies on an event selection via coincidence measurements.
The classic scheme
for the generation of continuous variable entanglement is the parametric
down-conversion. 
Starting with the first demonstration by Ou et al.~\cite{Ou:92},
the generation of entanglement in such systems has been achieved in 
several experiments~\cite{braunstein:05}.
It still remains, however, a challenge to generate entanglement in
macroscopic light rather than on the few photon level.
Promising candidates for the generation of macroscopic light entanglement
are optical amplifiers~\cite{morigi:06,morigi2:06,zhou:06,xiong:05,tan:05}. 
For example, it was shown recently
that a two-mode correlated spontaneous emission laser 
(CEL)~\cite{scully:85,scully:87} can lead to two-mode entanglement even when 
the average photon number in the field modes are very large~\cite{xiong:05,tan:05}.
In this setup, the gain medium can be thought of as a stream of suitably prepared 
atoms. 

From a conceptual point of view, a much simpler 
system relates to a single atom laser, where the gain medium is 
replaced by a single trapped atom. Such a  laser has recently 
been experimentally demonstrated by Kimble's group~\cite{mckeever:03}, 
where a single atom interacts with a single cavity mode. 
Thus the interesting question arises, whether a two-mode generalization 
of the single-atom laser also enables one to generate entanglement 
in macroscopic light. 

Therefore, here we consider a  single atom that interacts with two quantized modes 
of a doubly resonant cavity via two lasing transitions. 
In our model, the atomic level scheme is based on 
the single-atom laser experiment performed by Kimble's group~\cite{mckeever:03}, 
where dipole transitions between four hyperfine levels of atomic caesium were considered.  
In contrast to their experiment, we do not   work in the strong coupling 
regime since we are interested in the generation of large photon numbers. 
We show that, under certain realizable conditions, a two-mode single-atom laser 
can serve as a source of macroscopic entangled light. Macroscopic entanglement
can be achieved over a wide range of control parameters and initial
states of the cavity field.
 
An important technical question in the generation of continuous variable entanglement in
quantum optical systems is the way such entanglement can be measured
experimentally. This is a hotly discussed subject in recent years. Several 
inequalities involving the correlation of the field operators have been derived 
that are based on the separability condition of the field 
modes~\cite{simon:00,duan:00,shchukin:05,zubairy:06,agarwal:05,nha:06,guehne:06,giovannetti:03}.
A violation of these inequalities provides an
evidence of entanglement. These inequalities can, in general, provide only a 
sufficient  condition for entanglement and only, in some very specific
instances, lead to sufficient and necessary conditions for entanglement. In
this paper we use the inequality based on quadrature measurement of the field
variables for the test of entanglement. 

\section{MASTER EQUATION FOR THE DENSITY OPERATOR OF THE CAVITY MODES \label{model}}
\begin{figure}[t!]
\includegraphics[scale=1]{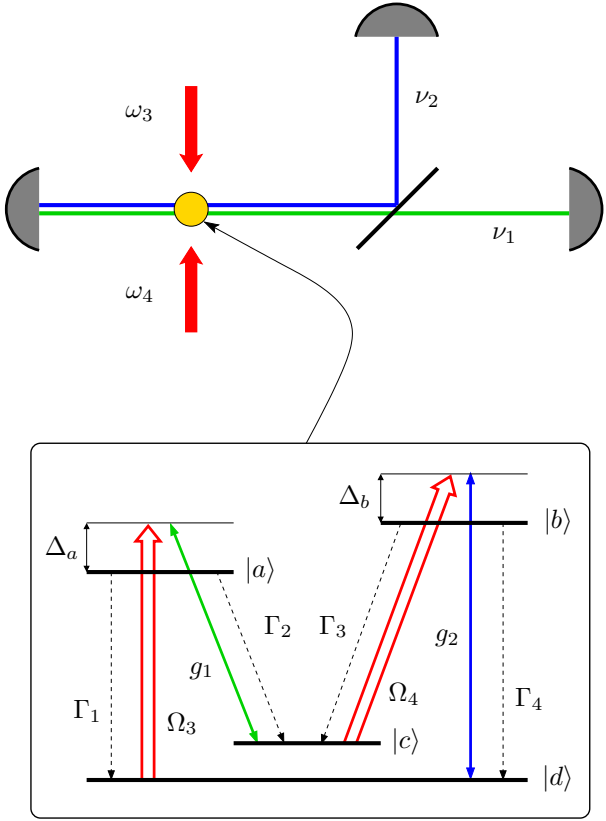}
\caption{\label{picture1} \small   
A single four-level atom is trapped in a doubly resonant cavity and 
interacts with two cavity modes and two classical laser fields. 
The inset shows the atomic level scheme.  
The  laser field with frequency $\omega_3$ and 
Rabi frequency $\Omega_3$ couples to the $\ket{a}\leftrightarrow\ket{d}$ 
transition, and the cavity mode with frequency $\nu_1$ and coupling constant 
$g_1$ interacts with the $\ket{a}\leftrightarrow\ket{c}$ 
transition. $\Delta_a$ is the detuning of the  fields $\Omega_3$ and $g_1$ with  
state $\ket{a}$.
The  laser field with frequency $\omega_4$ and 
Rabi frequency $\Omega_4$ drives  the $\ket{b}\leftrightarrow\ket{c}$ 
transition, and the second cavity mode with frequency $\nu_2$ and coupling constant 
$g_2$ interacts with the $\ket{b}\leftrightarrow\ket{d}$ 
transition. $\Delta_b$ is the detuning of the  fields $\Omega_4$ and $g_2$ with  
state $\ket{b}$. Spontaneous emission 
is denoted by dashed arrows, and the parameters 
$\Gamma_i$ are the  decay rates of the various transitions.  
  }
\end{figure}

We consider a single four-level atom trapped in a doubly resonant cavity (see Fig.~\ref{picture1}).  
The atom interacts with two (nondegenerate) cavity modes and two classical 
laser  fields. The intensities and frequencies of the two laser fields 
can be adjusted independently. The aim of this section is to derive 
an equation of motion for the reduced density operator $\rf$ of the 
two cavity modes.

We begin with a detailed description of the system shown in 
Fig.~\ref{picture1}. 
The first cavity mode with frequency $\nu_1$ couples to the atomic transition 
$\ket{a}\leftrightarrow\ket{c}$, and the second mode with frequency $\nu_2$ 
interacts with the atom on the $\ket{b}\leftrightarrow\ket{d}$ transition. 
In rotating-wave approximation (RWA), the interaction of the atom 
with the cavity modes  is described by the Hamiltonian 
\be
H_{\text{C}} =  \hbar  g_1 a_1 \ket{a}\bra{c} +  \hbar  g_2 a_2 \ket{b}\bra{d} + \text{H.c.} \,.
\ee 
%%%%%%%%%%%%%%%%%%%%%%%%%%%%%%%%%%%%
%%%%%%%%%%%%%%%%%%%%%%%%%%%%%%%%%%%%
Here $a_j$ ($a_j^{\dagger})$  is the annihilation (creation) operator of 
the cavity mode with frequency $\nu_j$ and coupling constant $g_j$ ($j\in\{1,2\}$). 
The detuning of the first cavity mode with the $\ket{a}\leftrightarrow\ket{c}$ 
transition is denoted by $\Delta_1$, and 
$\Delta_2$ is the detuning of the second mode with the  $\ket{b}\leftrightarrow\ket{d}$ 
transition, 
\be
\Delta_1 = \nu_1 - \omega_{ac}\, , \qquad \Delta_2=\nu_2 - \omega_{bd} \,. 
\ee
The resonance frequencies on the $\ket{a}\leftrightarrow\ket{c}$ and 
$\ket{b}\leftrightarrow\ket{d}$ transitions have been labeled by $\omega_{ac}$ 
and $\omega_{bd}$, respectively. 
%%%%%%%
In addition, the atom interacts with two classical laser fields. 
The first laser field with frequency $\omega_3$ and 
Rabi frequency $\Omega_3$ couples to the $\ket{a}\leftrightarrow\ket{d}$ 
transition, and the second field with frequency $\omega_4$ and 
Rabi frequency $\Omega_4$ coherently drives the $\ket{b}\leftrightarrow\ket{c}$ 
transition. In rotating-wave approximation, 
the atom-laser interaction reads 
\be
H_{\text{L}} =  -\hbar \Omega_3\ket{a}\bra{d} e^{-i\omega_3 t}
 -\hbar\Omega_4\ket{b}\bra{c} e^{-i\omega_4 t} + \text{H.c.}  \,. 
\ee
Note that the Rabi frequencies $\Omega_3=|\Omega_3|\exp(i\phi_3)$ 
and $\Omega_4 = |\Omega_4|\exp(i\phi_4)$ are complex 
numbers, and $\phi_3$ and $\phi_4$ are determined by the phase of 
the laser fields. 
The detuning of the  laser fields with the corresponding  atomic transitions are 
\be
\Delta_3 = \omega_3 -\omega_{ad}\,,\quad \Delta_4=\omega_4 - \omega_{bc}\, ,
\ee
where $\omega_{ad}$ and $\omega_{bc}$ are the resonance frequencies 
on the $\ket{a}\leftrightarrow\ket{d}$ and $\ket{b}\leftrightarrow\ket{c}$ 
transitions, respectively.  

The free time evolution of the cavity modes is governed by 
\be 
H_{\text{R}} =  \hbar\nu_1 a_1^{\dagger}a_1+ \hbar \nu_2 a_2^{\dagger}a_2 \,,
\ee
and $H_{\text{A}}$ is the  free Hamiltonian  of the atomic degrees of freedom, 
\be
H_{\text{A}} =  \hbar\omega_a \ket{a}\bra{a} + \hbar\omega_b \ket{b}\bra{b} + \hbar\omega_c \ket{c}\bra{c}
+ \hbar\omega_d \ket{d}\bra{d} \,.
\ee
%%%%%%%%%%%%%%%%%%%%%%
With these definitions, we arrive at the master equation 
for the combined system of the atomic degrees 
of freedom  and the two cavity modes, 
\be
\dot{\vro} = - \frac{i}{\hbar} [ H_{\text{R}}  + H_{\text{A}}  + H_{\text{L}} + H_{\text{C}}, \vro ] +\mc{L}_{\gamma}\vro\,.
\label{master_eq}
\ee
%%%%%%%%%%%%%%%%%%%%%%%%%%%%
The last term in Eq.~(\ref{master_eq}) accounts for spontaneous emission 
and is given by  
\be 
\mc{L}_{\gamma}\vro = -\frac{1}{2} \sum\limits_{i=1}^4 \Gamma_i 
\left( S_i^+ S_i^- \vro  + \vro S_i^+S_i^-   - 2 S_i^-\vro S_i^+\right) \,,
\ee 
where the atomic transition operators are defined as 
\begin{align}
S_1^+ =\ket{a}\bra{d} & ,\quad S_2^+ = \ket{a}\bra{c} \, ,\notag \\
S_3^+ =\ket{b}\bra{c} & ,\quad S_4^+ = \ket{b}\bra{d}\,,\quad S_i^- = (S_i^+)^{\dagger}\,.
\end{align}
The parameters $\Gamma_i$ are the decay rates of the various atomic transitions 
(see  Fig.~\ref{picture1}). 
%%%%%%%%%%%%%%%%%%%%%%%%%%

In a next step, we derive from Eq.~(\ref{master_eq}) 
the master equation for the density 
operator $\rf$ of the cavity modes, 
\be
\rf= \text{Tr}_{\text{A}}\vro = 
\vro_{aa}+\vro_{bb}+\vro_{cc}+\vro_{dd}\,,
\ee
and $\vro_{\nu\nu}$ denotes $\bra{\nu}\vro\ket{\nu}$. 
To this end, we apply a unitary  transformation $W=W_{\text{R}} \otimes W_{\text{A}}$ to 
Eq.~(\ref{master_eq}), where $W_{\text{R}} =  \exp[i H_{\text{R}} t/\hbar]$  acts only on 
the cavity modes, and 
\begin{align}
W_{\text{A}} = & \exp[i(H_{\text{A}} + \hbar \Delta_3 \ket{a}\bra{a} + \hbar\Delta_4\ket{b}\bra{b}) t/\hbar]
\end{align}
acts only on the atomic degrees of freedom. 
As indicated in Fig.~\ref{picture1}, 
we assume that the condition of two-photon resonance is fulfilled, i.e. 
\be
\Delta_a = \Delta_1 = \Delta_3\,,\quad\Delta_b =  \Delta_2 = \Delta_4 \,.
\label{twophoton}
\ee
The density operator in the new frame is 
denoted by $\tilde{\vro} = W\vro W^{\dagger}$ and 
obeys the equation of motion
\be
\dot{\tilde{\vro}} = - \frac{i}{\hbar} [ H_0 + H_{\text{C}},\tilde{\vro} ] +\mc{L}_{\gamma}\tilde{\vro}\,,
\label{master_eq2}
\ee
where 
\begin{align}
H_0 = & -\hbar \Delta_a \ket{a}\bra{a}-\hbar\Delta_b \ket{b}\bra{b}  \notag \\
& -\hbar\left(\Omega_3\ket{a}\bra{d}
 + \Omega_4\ket{b}\bra{c}  + \text{H.c.}\right)  \,.
\end{align}
The two-photon condition Eq.~(\ref{twophoton}) 
ensures that the Hamiltonian $H_0 + H_{\text{C}}$ in  Eq.~(\ref{master_eq2}) is 
time-independent. 
The master equation for the transformed density operator 
$\tilde{\vro}_{\text{F}}$ of the cavity modes  is obtained if 
we trace over the atomic degrees of freedom in Eq.~(\ref{master_eq2}), 
\be
\dot{\tilde{\vro}}_{\text{F}}  =  -i  g_1  [ a_1^{\dagger}, \tilde{\vro}_{ac} ]   -i g_2  [ a_2^{\dagger}, \tilde{\vro}_{bd}  ]   
+\text{H.c.}\,.
\label{master_field_W}
\ee
In order to eliminate the  coherences $\tilde{\vro}_{ac}$ and $\tilde{\vro}_{bd}$ 
from Eq.~(\ref{master_field_W}), we apply the standard methods 
of laser theory (see, e.g., Chapter 14 in~\cite{zubairy:qo}). 
We restrict the analysis to the linear theory and solve Eq.~(\ref{master_eq2}) 
to first order in the coupling constants $g_1$ and $g_2$. To this end, we  expand  
the density operator $\rft$ in Eq.~(\ref{master_eq2}) 
as $\tilde{\vro} = \vro_0 +\vro_C$ and retain only terms up to first order 
with respect to $H_{\text{C}}$. This procedure yields  
two uncoupled equations for $\vro_0$ and $\vro_C$,
\begin{align}
\dot{\vro_0} = & \mc{L}_0 \vro_0 \,, \label{rho_0} \\
\dot{\vro_C} = & \mc{L}_0 \vro_C -\frac{i}{\hbar}[H_{\text{C}} ,\vro_0] \,,
\label{perturbation}
\end{align}
and  the superoperator $\mc{L}_0$ is defined as 
\be
\mc{L}_0(\cdot) = -  \frac{i}{\hbar} [ H_0 , \,\cdot\, ] +\mc{L}_{\gamma}(\cdot)\,. 
\ee
Here the centered dot denotes the position of the argument of  $\mc{L}_0$. 
The zeroth-order equation~(\ref{rho_0}) describes  the interaction of 
the atom with the classical laser fields to all orders, and Eq.~(\ref{perturbation}) is 
the first-order equation.  
%%%%%%%%%%%%%%%%%%%%%%%%%%%%%%%%%%
The steady state solution for  $\tilde{\vro}_{ac}$ and $\tilde{\vro}_{bd}$ 
can be obtained  if the steady-state solution   for $\vro_0$ 
is plugged in Eq.~(\ref{perturbation}).  We find 
\begin{align}
i g_1 \tilde{\vro}_{ac}  = \alpha_{11} a_1 \rft + \alpha_{12} a_2^{\dagger} \rft
 + \beta_{11} \rft a_1  + \beta_{12} \rft a_2^{\dagger} \, , \notag \\[0.2cm]
i g_2 \tilde{\vro}_{bd}  = \alpha_{22} a_2 \rft + \alpha_{21} a_1^{\dagger} \rft
 + \beta_{22} \rft a_2  + \beta_{21} \rft a_1^{\dagger} \, ,
 \label{coherences}
\end{align}
and the coefficients $\alpha_{ij}$ and $\beta_{ij}$ are defined in  Appendix~\ref{coefficients}. 
Next we substitute Eq.~(\ref{coherences}) in Eq.~(\ref{master_field_W}) to 
obtain the equation of motion for $\tilde{\vro}_{\text{F}}$. 
Finally, we transform 
$\tilde{\vro}_{\text{F}}$ back with respect to $W_R$ and obtain the 
equation of motion for the density operator $\rf$ of the cavity modes,
\begin{widetext}
\begin{align}
\dot{\rf} = 
& - i\,\nu_1 [ a_1^{\dagger}a_1 , \rf ] - i\,\nu_2 [ a_2^{\dagger}a_2 , \rf ] \label{master_field} \\[0.2cm]
& -\left[ \alpha_{11} a_1^{\dagger} a_1 \rf 
+ \alpha_{11}^* \rf a_1^{\dagger} a_1 
- (\alpha_{11}+\alpha_{11}^*) a_1 \rf a_1^{\dagger} -\beta_{11}^* a_1 a_1^{\dagger} \rf 
- \beta_{11} \rf a_1 a_1^{\dagger} 
+ (\beta_{11}+\beta_{11}^* ) a_1^{\dagger} \rf a_1 \right] \notag \\[0.2cm]
 & -\left[ \alpha_{22} a_2^{\dagger} a_2 \rf + \alpha_{22}^* \rf a_2^{\dagger} a_2 
- (\alpha_{22}+\alpha_{22}^*) a_2 \rf a_2^{\dagger} -\beta_{22}^* a_2 a_2^{\dagger} \rf 
- \beta_{22} \rf a_2 a_2^{\dagger} 
+ (\beta_{22}+\beta_{22}^* ) a_2^{\dagger} \rf a_2 \right] \notag \\[0.2cm]
& -\left[  (\alpha_{12} + \alpha_{21} ) a_1^{\dagger} a_2^{\dagger} \rf -(\beta_{12}
+\beta_{21})\rf a_1^{\dagger} a_2^{\dagger} 
-(\alpha_{21} - \beta_{12}) a_1^{\dagger} \rf a_2^{\dagger} 
-(\alpha_{12} -\beta_{21}) a_2^{\dagger}\rf a_1^{\dagger}              \right]\, \exp[-i(\nu_1 + \nu_2) t]\notag \\[0.2cm]
& -\Big[  (\alpha_{12}^* + \alpha_{21}^* ) \rf a_1 a_2   -(\beta_{12}^* +\beta_{21}^*)  a_1  a_2 \rf  
-(\alpha_{21}^* - \beta_{12}^* ) a_2 \rf a_1  
-(\alpha_{12}^* -\beta_{21}^* ) a_1 \rf a_2        \Big]\,\exp[i(\nu_1 + \nu_2) t]\notag \\[0.2cm]
 &-\kappa_1 \left( a_1^{\dagger} a_1 \rf +\rf a_1^{\dagger} a_1 -2 a_1 \rf a_1^{\dagger}\right)
-\kappa_2 \left( a_2^{\dagger} a_2 \rf +\rf a_2^{\dagger} a_2 -2 a_2 \rf a_2^{\dagger}\right)\,. \notag
\end{align}
\end{widetext}
In the last line of Eq.~(\ref{master_field}), we included the damping of the cavity field. 
The damping constants of the cavity modes are denoted by  $\kappa_1$ and 
$\kappa_2$, respectively.  

In the master equation~(\ref{master_field}), the two classical laser fields are taken into 
account to all orders in the Rabi frequencies $\Omega_3$ and $\Omega_4$. 
On the contrary, the two quantum fields 
inside the cavity are only treated to second order in the coupling 
constants $g_1$ and $g_2$.  This approximation means that we 
ignore saturation effects and operate in the regime of linear amplification. 
It is justified  if the Rabi frequencies associated with the quantum fields 
are small as compared to other system parameters which dominate 
the time evolution. 

\section{ENTANGLEMENT OF THE CAVITY FIELD \label{results}}
In this Section we show that the system depicted in Fig.~\ref{picture1} 
can serve as a source of macroscopic entangled light. 
We employ the sufficient inseparability criterion derived in~\cite{duan:00}  
to provide evidence for the entanglement of  the two field modes.  

By definition, the quantum state $\rf$ of the cavity field is said to 
be entangled if and only if  it is nonseparable, and 
$\rf$  is separable if 
and only if it can be written as
\be
\rf = \sum\limits_j p_j \vro_j^{(1)} \otimes \vro_j^{(2)}\,. 
\ee
Here  $\vro_j^{(1)}$ and $\vro_j^{(2)}$ are  normalized states 
of the  modes 1 and 2, respectively, and the parameters  
$p_j\ge 0$ comply with $\sum_j p_j =1$.  
The criterion derived in~\cite{duan:00} 
states that the system is in an entangled quantum state if   
the total variance of two Einstein-Podolsky-Rosen (EPR) type   
operators $\hat{u}$ and $\hat{v}$ of the two modes satisfy the inequality   
\be
\meanb{\left(\Delta\hat{u}\right)^2  + \left(\Delta\hat{v}\right)^2} < 2 \,,
\label{criterion}
\ee
where 
\be
\hat{u} = \hat{x}_1 + \hat{x}_2 \,,\qquad \hat{v} = \hat{p}_1 - \hat{p}_2\,.
\label{u_and_v}
\ee
Here $\hat{x}_k$ and $\hat{p}_k$ are local operators which 
correspond to mode $k$ with frequency $\nu_k$. 
They must obey the commutation relation
\be
[\hat{x}_k, \hat{p}_l] = i\delta_{kl}\,, 
\ee
but are otherwise arbitrary. For the physical system considered here, 
it turns out that the following  quadrature operators 
\be
\hat{x}_k = (b_k + b_k^{\dagger})/\sqrt{2} \quad\text{and}\quad \hat{p}_k = (b_k - b_k^{\dagger})/(\sqrt{2} i) 
\label{quadratures}
\ee
are the best choice, where 
\be
b_k(t) =a_k \exp[i\nu_k t]\quad \text{and}\quad b_k^{\dagger}(t) =a_k^{\dagger} \exp[-i\nu_k t] \,.
\ee
With the help of Eqs.~(\ref{u_and_v}) and~(\ref{quadratures}), we express 
the total variance of the operators $\hat{u}$ and $\hat{v}$ in terms of 
the operators $b_k$ and $b_k^{\dagger}$, 
\begin{align}
& \meanb{\left(\Delta\hat{u}\right)^2  + \left(\Delta\hat{v}\right)^2}  =  
2\left[
1 + \mean{b_1^{\dagger}b_1} + \mean{b_2^{\dagger}b_2} 
+ \mean{b_1 b_2} \right. \notag \\
& \left. + \mean{b_1^{\dagger} b_2^{\dagger}}  - \mean{b_1}\mean{b_1^{\dagger}}  - \mean{b_2}\mean{b_2^{\dagger}}
 - \mean{b_1}\mean{b_2} - \mean{b_1^{\dagger}}\mean{b_2^{\dagger}}
\right]   \,. 
\label{lhs_criterion}
\end{align}
In Appendix~\ref{meanvalues}, we outline the calculation of 
the mean values that enter Eq.~(\ref{lhs_criterion}). 

Next we classify several parameter regimes for which  the 
inequality~(\ref{criterion}) is fulfilled. 
In a first step, we consider the case where the  
Rabi frequency $|\Omega_3|$ and the detuning $\Delta_b$  
are much larger than the parameters $|\Delta_a|,\,|\Omega_4|,\,\Gamma_i$  
$(i\in\{1,2,3,4\})$, i.e.  
\be  
|\Omega_3|,\, |\Delta_b| \gg |\Delta_a|,\,|\Omega_4|,\,\Gamma_i \,. 
\label{parametric_condition}
\ee
%%%%%%%%%%%%%%%%%%%%%%%%%%%%%%%%%%%%
\begin{figure}[t!]
\includegraphics[scale=1]{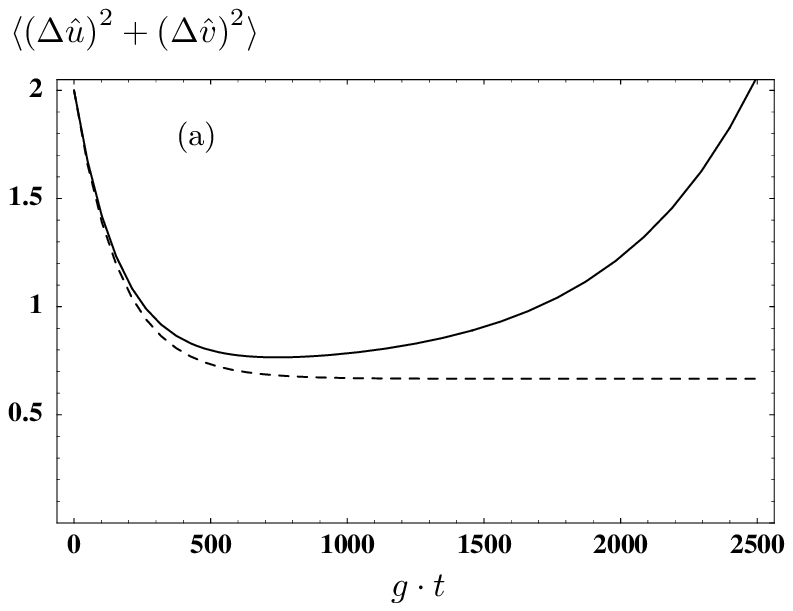}
\includegraphics[scale=1]{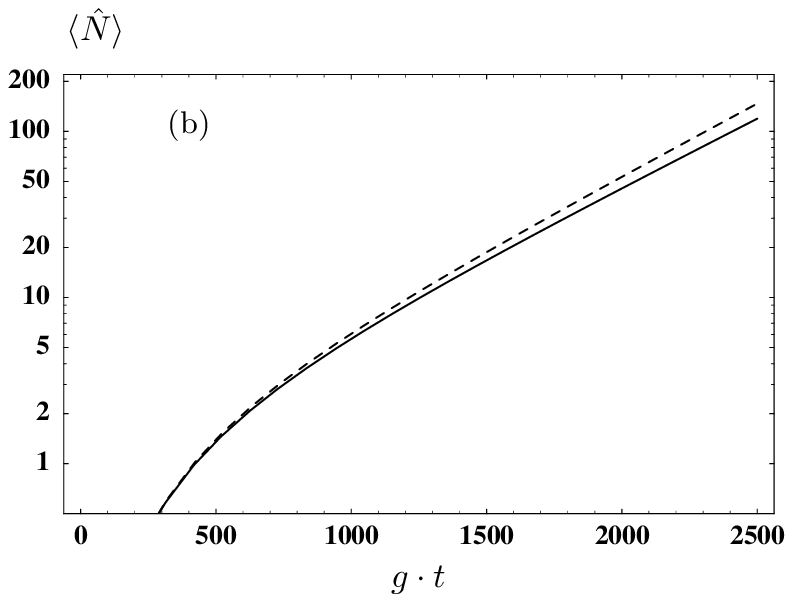}
\caption{\label{picture2} \small 
(a) Time evolution of  
$\mean{\left(\Delta\hat{u}\right)^2 + \left(\Delta\hat{v}\right)^2}$. 
The mean value of the total number of photons $\mean{\hat{N}}$ 
is shown in (b) on a logarithmic scale. 
At $t=0$, the cavity field is assumed to be in the vacuum state. 
The dashed curves  were obtained with the density operator of the parametric oscillator in 
Eq.~(\ref{master_parametric}), and the solid curves  correspond 
to the full density operator in Eq.~(\ref{master_field}). 
The parameters are 
$g_1 = g_2 =g$, 
$|\Omega_3| = 25g$, $|\Omega_4| = 2 g$, $\Gamma_1=\Gamma_2=\Gamma_3=\Gamma_4=5g$, 
$\Delta_a = 0$, $\Delta_b =40 g$, $\kappa_1=\kappa_2=10^{-3}g$ and $\phi_3 + \phi_4 = \pi/2$. 
}
\end{figure}
%%%%%%%%%%%%%%%%%%%%%%%%%%%%%%%%%%%%
If these conditions are fulfilled, the parameters $\alpha_{ij}$ and $\beta_{ij}$ 
in Eqs.~(\ref{a11})-(\ref{b21}) of Appendix~\ref{coefficients}  reduce to 
\begin{align}
\alpha_{11} & \approx 0, \,\alpha_{22} \approx 0, \, \beta_{11} \approx 0, \, \beta_{22} \approx 0\,, \notag \\[0.2cm]
\alpha_{21} & \approx 0, \, \beta_{12} \approx  0, \, \alpha_{12}  \approx \beta_{21} 
 \approx - i \alpha \exp[i(\phi_3+ \phi_4)t]\,,  \notag \\[0.2cm]
\alpha  & =  g_1 g_2 \frac{|\Omega_4|}{|\Omega_3| \Delta_b}\, . 
\label{parametric_para}
\end{align}
In these equations, $\phi_3$ and $\phi_4$ are  the phases of the  
classical laser fields with Rabi frequencies 
$\Omega_3=|\Omega_3|\exp(i\phi_3)$   and $ |\Omega_4|\exp(i\phi_4) $, respectively (see Sec.~\ref{model}). 
If the approximate parameters in Eq.~(\ref{parametric_para}) are plugged 
in Eq.~(\ref{master_field}), we obtain the equation of motion for the 
density operator $\rf$ of the cavity modes in the limit~(\ref{parametric_condition}),
\begin{align}
\dot{\rf} = 
& - i\,\nu_1 [ a_1^{\dagger} a_1 , \rf ] - i\,\nu_2 [ a_2^{\dagger} a_2 , \rf ] + i [H_\text{P}, \rf]   \notag \\[0.2cm]
&-\kappa  \left( a_1^{\dagger} a_1 \rf +\rf a_1^{\dagger} a_1 -2 a_1 \rf a_1^{\dagger}\right. \notag \\[0.2cm]
&  \hspace*{1cm} \left. +  a_2^{\dagger} a_2 \rf +\rf a_2^{\dagger} a_2 -2 a_2 \rf a_2^{\dagger}\right)\,,
\label{master_parametric}
\end{align}
where 
\begin{align}
H_\text{P} = & \alpha   a_1^{\dagger}a_2^{\dagger}  \exp[i(\phi_3+ \phi_4)t] \exp[-i(\nu_1+\nu_2)t] \notag \\[0.2cm] 
& + \alpha  a_1 a_2   \exp[-i(\phi_3+ \phi_4)t] \exp[i(\nu_1+\nu_2)t] \,.
\label{H_parametric}
\end{align}
Here we assumed for the sake of simplicity that 
the decay rates of the cavity modes are equal, $\kappa_1 = \kappa_2=\kappa$. 
We identify Eq.~(\ref{master_parametric}) as  the master equation for a 
nondegenerate parametric oscillator in the parametric approximation~\cite{zubairy:qo}.  
Note that this parametric limit was also obtained in the case of a 
two-mode correlated spontaneous emission laser 
discussed in~\cite{xiong:05}. 
%%%%%%%%%%%%%%%%%%%%%%%%%%%%%%%%%%%
Next we evaluate the total variance of the operators 
$\hat{u}$ and $\hat{v}$  in Eq.~(\ref{lhs_criterion})  and  the 
mean number of photons 
$\mean{\hat{N}} = \mean{a_1^{\dagger}a_1 + a_2^{\dagger}a_2} =  \mean{b_1^{\dagger}b_1 + b_2^{\dagger}b_2}$  
with the approximate density operator $\rf$ in Eq.~(\ref{master_parametric}).       
If the sum of the laser phases obeys  $\phi_3 + \phi_4 = \pi/2$,  we obtain~\cite{xiong:05} 
\begin{align}
  &  \hspace*{-0.5cm} \meanb{\left(\Delta\hat{u}\right)^2 + \left(\Delta\hat{v}\right)^2}(t) =  
\left[\meanb{\left(\Delta\hat{u}\right)^2 + \left(\Delta\hat{v}\right)^2}(0)  \right. \notag \\[0.2cm]
& \hspace*{3.1cm} \left. -\frac{2\kappa}{\alpha + \kappa}\right]e^{-2 (\alpha + \kappa)t} + \frac{2\kappa}{\alpha + \kappa} \,, 
\label{simple_ent} \\[0.2cm]
&  \hspace*{-0.5cm} \meanb{\hat{N}}(t) = \left[  \meanb{\hat{N}}(0)  - \frac{\alpha^2}{\kappa^2 - \alpha^2}\right]
 \cosh(2\alpha t) e^{-2\kappa t}  \notag \\[0.2cm]
& \hspace*{1cm} - \left[ \frac{\alpha \kappa}{\kappa^2 -\alpha^2} + \meanb{a_1 a_2 + a_1^{\dagger} a_2^{\dagger}}(0)\right] 
\sinh(2\alpha t) e^{-2\kappa t}   \notag \\[0.2cm]
& \hspace*{1cm} + \frac{\alpha^2 }{\kappa^2 -\alpha^2} \,. 
\label{simple_num}
\end{align}
%%%%%%%%%%%%%%
% discussion
%%%%%%%%%%%%%%
It follows from Eq.~(\ref{simple_ent})  
that the entanglement criterion  in Eq.~(\ref{criterion}) 
is satisfied for any initial state of the cavity field if $(\alpha + \kappa) t \gg 1$ and $\alpha>0$~\cite{xiong:05} .   
The time evolution of the total variance of the operators $\hat{u}$ and $\hat{v}$ 
is shown in Fig.~\ref{picture2}(a).  The dashed curve shows 
$\mean{\left(\Delta\hat{u}\right)^2  + \left(\Delta\hat{v}\right)^2}$ according to 
 Eq.~(\ref{simple_ent}), and the solid line corresponds to the general case where the 
mean values in Eq.~(\ref{lhs_criterion})  are  evaluated with the full density operator 
$\rf$ in Eq.~(\ref{master_field}). 
The cavity modes are assumed to be in the vacuum state initially, and the 
parameters comply with condition~(\ref{parametric_condition}).  
It follows from Fig.~\ref{picture2} that the approximate 
result in Eq.~(\ref{simple_ent})  is only in good agreement with the 
exact solution if $g t  < 300$.  While the light field 
remains in an entangled state in the parametric case, 
the exact solution demonstrates that the entanglement of 
the cavity field exists only for  a finite period of time.  

Next we discuss the time evolution of the mean number 
of photons $\mean{\hat{N}}$.  
According to  Eq.~(\ref{simple_num}),  $\mean{\hat{N}}$ 
grows exponentially with time 
for any initial state of the cavity field, provided that 
$(\alpha-\kappa)t \gg 1$ and $\alpha > \kappa$~\cite{xiong:05}.  
The time evolution of $\mean{\hat{N}}$  is shown in 
Fig.~\ref{picture2}(b) on a logarithmic scale.   
In contrast to 
$\mean{\left(\Delta\hat{u}\right)^2  + \left(\Delta\hat{v}\right)^2}$, 
the result for $\mean{\hat{N}}$ in  the parametric approximation (dashed line) 
is in good agreement with the exact solution (solid line) even for $g t \gg 300$.  
Moreover,  Fig.~\ref{picture2}(b) shows that   
the mean number of photons  grows exponentially if the scaled time $g t$ 
is sufficiently large. 
%%%%%%%%%%%%%%%%%%%%%%%%%%%%%%%%%%%%%%%%%%
\begin{figure}[t!]
\includegraphics[scale=1]{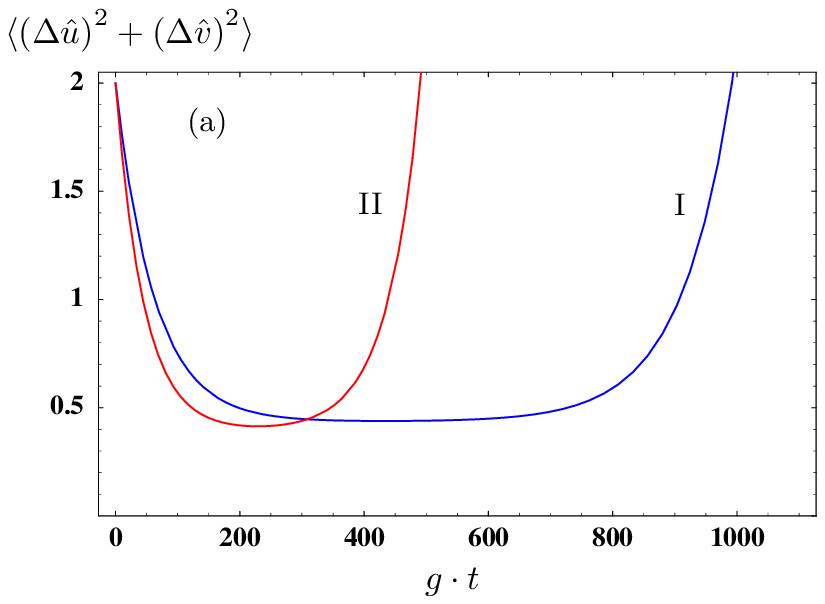}
\includegraphics[scale=1]{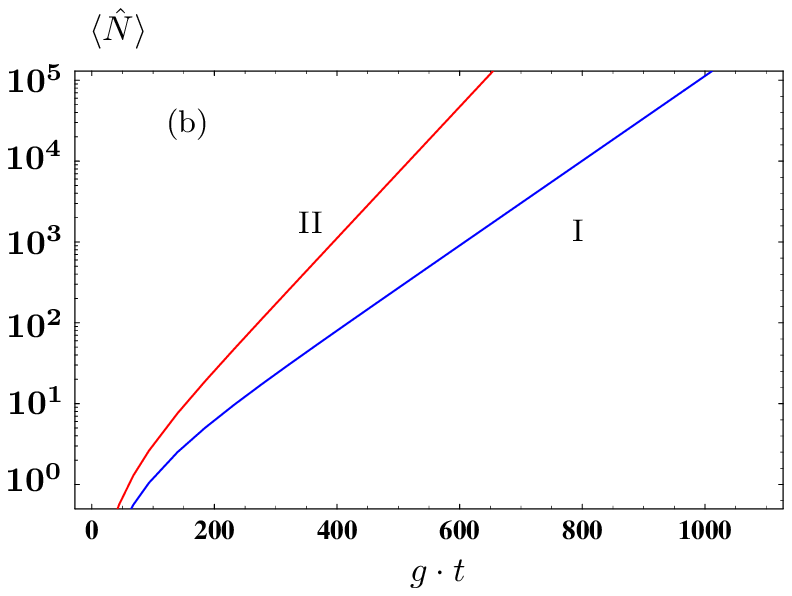}
\caption{\label{picture3} \small (Color online) 
(a) Time evolution of   
$\mean{\left(\Delta\hat{u}\right)^2  + \left(\Delta\hat{v}\right)^2}$. 
The mean value of the total number of photons $\mean{\hat{N}}$ 
is shown in (b) on a logarithmic scale. 
At $t=0$, the cavity field is assumed to be in the vacuum state, and we set 
$\Gamma_1=\Gamma_2=\Gamma_3=\Gamma_4=5g$, 
$g_1 = g_2 =g$, 
 $\kappa_1=\kappa_2=10^{-3}g$ and $\phi_3 + \phi_4 = \pi/2$. 
The parameters for the curves labeled with I are 
$|\Omega_3| = 25g$, $|\Omega_4| = 9.8 g$, 
$\Delta_a = 0$, $\Delta_b =43 g$, and for II we have 
$|\Omega_3| = 15g$, $|\Omega_4| = 6 g$, 
$\Delta_a = 0$, $\Delta_b =32.5 g$. 
}
\end{figure}
%%%%%%%%%%%%%%%%%%%%%%%%%%%%%%%%%%%%%%%%%%%%%%%%%

%%%%%%%%%%%%%%%%%%%%%%
% intuitive picture
%%%%%%%%%%%%%%%%%%%%%%
According to Fig.~\ref{picture2},  the entangled state of the 
cavity field contains up to 
$\mean{\hat{N}}\approx 110$ photons on average. 
It follows  that the single-atom laser 
depicted in Fig.~\ref{picture1} can give 
rise to an entangled quantum state of the 
two cavity modes if the parameters are in 
agreement with condition~(\ref{parametric_condition}). 
If this condition holds, level $\ket{b}$ is almost not 
excited due to the large detuning $\Delta_b$, and  
states $\ket{c}$ and $\ket{d}$ are coupled 
via a two-photon process. In contrast, the transitions 
$\ket{d}\leftrightarrow\ket{a}$ and $\ket{c}\leftrightarrow\ket{a}$ 
are driven resonantly.  
In this situation, the structure of the Hamiltonian $H_{\text{P}}$ in 
Eq.~(\ref{H_parametric}) implies that the system can only emit 
photons into the cavity fields  in pairs, where one photon is emitted in mode 1  and the other photon in mode 2. 
If the cavity field is initially in the vacuum state $\ket{0,0}$, it will evolve under 
the influence of $H_{\text{P}}$ into the entangled state
\be 
a\,\ket{0,0} + b\,\ket{1,1} + c\,\ket{2,2} +\ldots \,,
\ee
where $a$, $b$ and $c$ are complex coefficients.  
If the complicated master equation (\ref{master_field}) 
can be reduced under certain conditions  to the parametric equation (\ref{master_parametric}), it is thus 
clear that  a macroscopic entangled state is generated. 
%%%%%%%%%%%%%%%%%%%%%%%%%%%%%%%%

Due to the symmetry in the atomic level scheme, it 
is possible to reverse the role of the transitions 
$\ket{d}\leftrightarrow\ket{a}\leftrightarrow\ket{c}$ and 
$\ket{c}\leftrightarrow\ket{b}\leftrightarrow\ket{d}$.   
In this case, the  detuning $\Delta_a$ is large and 
the transitions $\ket{d}\leftrightarrow\ket{b}$ and 
$\ket{c}\leftrightarrow\ket{b}$ are driven resonantly.  
Condition~(\ref{parametric_condition}) then has 
to be replaced by 
\be 
|\Omega_4|,\, |\Delta_a| \gg |\Delta_b|,\,|\Omega_3|,\,\Gamma_i \,, 
\label{parametric_condition2}
\ee 
and the only nonvanishing coefficients in Eq.~(\ref{master_field}) 
are now determined by 
 $\alpha_{21}  \approx \beta_{12} 
 \approx - i \alpha' \exp[i(\phi_3+ \phi_4)t]$, where 
$\alpha' =g_1 g_2 |\Omega_3|/(|\Omega_4| \Delta_a)$. 
It follows that the results in Eqs.~(\ref{master_parametric}),~(\ref{simple_ent})  
and~(\ref{simple_num})  are also  valid if condition~(\ref{parametric_condition2})  
holds, provided that $\alpha$ is replaced by  $\alpha'$.

%%%%%%%%%%%%%%%%%%%%%%
% beyond parametric approximation
%%%%%%%%%%%%%%%%%%%%%%
We now demonstrate that it can be advantageous to 
consider parameters which do not comply with 
conditions~(\ref{parametric_condition}) 
or~(\ref{parametric_condition2}).  
Since the approximate results in Eqs.~(\ref{simple_ent}) and~(\ref{simple_num})  
do not apply in this case, we evaluate the mean values 
$\mean{\left(\Delta\hat{u}\right)^2  + \left(\Delta\hat{v}\right)^2}$  
and $\mean{\hat{N}}$ only with the exact density 
operator $\rf$ in Eq.~(\ref{master_field}). 
The time evolution of  
$\mean{\left(\Delta\hat{u}\right)^2  + \left(\Delta\hat{v}\right)^2}$  
is shown in Fig.~\ref{picture3}(a) for two sets of parameters. 
As compared to the parameters chosen for Fig.~(\ref{picture2}), 
the magnitude of the  Rabi frequency $\Omega_4$ has been 
increased such that $|\Omega_3|$ is still larger, but not much 
larger than $|\Omega_4|$.  It follows from Fig.~\ref{picture3}(a) 
that the entanglement criterion in Eq.~(\ref{criterion}) 
is fulfilled for shorter times as compared to the solid line 
in Fig.~\ref{picture2}(a).  
On the other hand, Fig.~\ref{picture3}(b) shows 
that  the mean  number of photons can be much 
larger as compared to Fig.~\ref{picture2}(b). 
For curve I of Fig.~\ref{picture3}(a), the maximum mean number of 
photons for which the entanglement criterion~(\ref{criterion}) 
is still fulfilled is $\mean{\hat{N}}\approx 10.2 \times 10^{4}$.  
The same number for the parameters of curve II reads 
$\mean{\hat{N}}\approx 6100$.   As compared to Fig.~\ref{picture2}, 
the maximum mean number of photons can be   enhanced by 
several orders of magnitude.  
%%%%%%%%%%%%%%%%%%%%%%%%%%%%%%%%%%%%%
\begin{figure}[t!]
\includegraphics[scale=1]{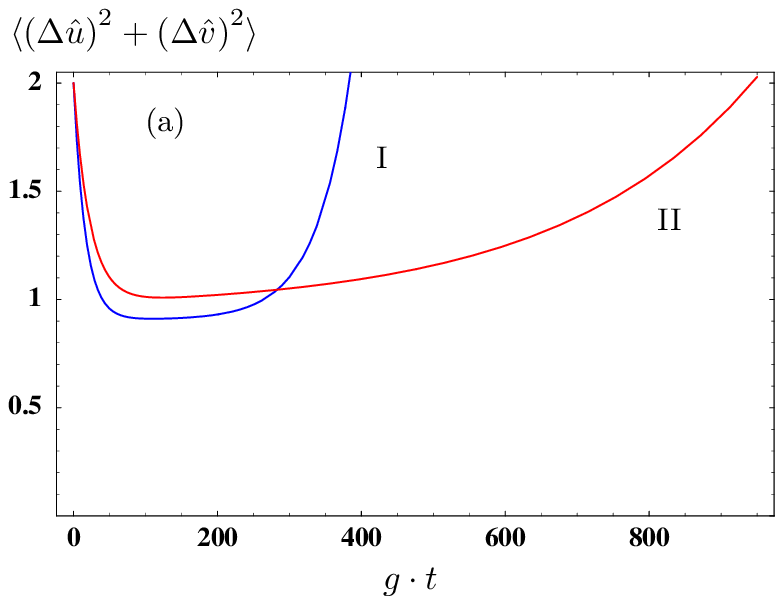}
\includegraphics[scale=1]{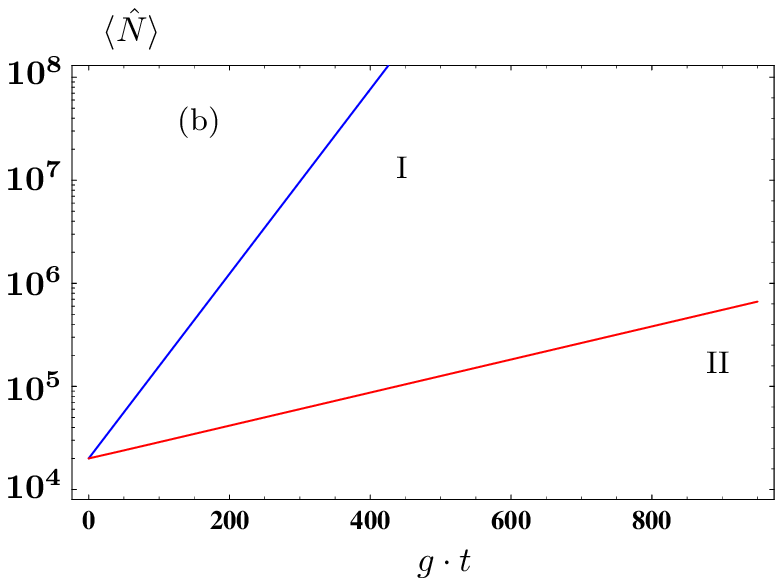}
\caption{\label{picture4} \small (Color online) 
(a) Time evolution of   
$\mean{\left(\Delta\hat{u}\right)^2  + \left(\Delta\hat{v}\right)^2}$. 
The mean value of the total number of photons $\mean{\hat{N}}$ 
is shown in (b) on a logarithmic scale. 
At $t=0$, the cavity field is assumed to be in the coherent state 
$\ket{100,-100}$, and we set 
$\Gamma_1=\Gamma_2=\Gamma_3=\Gamma_4=2g$, 
$g_1 = g_2 =g$, 
 $\kappa_1=\kappa_2=10^{-2}g$ and $\phi_3 + \phi_4 = \pi/2$. 
The parameters for the curves labeled with I are 
$|\Omega_3| = 10g$, $|\Omega_4| = 5 g$, 
$\Delta_a = 0$, $\Delta_b =15 g$, and for II we have 
$|\Omega_3| = 10g$, $|\Omega_4| =2 g$, 
$\Delta_a = 0$, $\Delta_b =15 g$. 
}
\end{figure}
%%%%%%%%%%%%%%%%%%%%%%%%%%%%%%%%%%%%

Finally, we consider the case where the quantum state of the cavity field 
is initially the coherent state $\ket{100,-100}$.  
The time evolution of 
$\mean{\left(\Delta\hat{u}\right)^2  + \left(\Delta\hat{v}\right)^2}$  
and $\mean{\hat{N}}$ is shown in Fig.~\ref{picture4} 
for two sets of parameters. All mean values were evaluated with 
the exact density operator in Eq.~(\ref{master_field}). 
For curve I, the magnitude of the Rabi 
frequency $\Omega_4$  is larger as compared 
to  curve II. All other parameters are the same for curve I and II. 
It can be seen from Fig.~\ref{picture4}(a) 
that the entanglement criterion is fulfilled for shorter  times 
if $|\Omega_4|$ is increased. 
In contrast, the mean number of photons can be greatly 
enhanced if the value of $|\Omega_4|$ is increased, 
as can be seen from Fig.~\ref{picture4}(b).  
Similar conclusions can be drawn from Fig.~\ref{picture3}, where 
the initial state of the cavity field is the vacuum. 
The comparison of Figs.~\ref{picture3} and~\ref{picture4} 
shows that the mean number of photons can be much 
larger than in Fig.~\ref{picture3} if the cavity field is 
initially prepared in a coherent state. 
Due to the large mean number of photons in the cavity modes, 
the system may leave the regime of 
linear amplification such that saturation effects   
modify the curve progression in Fig.~\ref{picture4}.  These 
effects are described by terms that go beyond the second-order 
expansion of the atom-cavity coupling and are neglected here. 
According to  the linear theory, the maximum mean number of 
photons for which the entanglement criterion~(\ref{criterion}) 
is still fulfilled is  $\mean{\hat{N}}\approx 6.5 \times 10^{5}$ 
in the case of curve II of Fig.~\ref{picture4}, and 
in the case of curve I the 
entangled cavity field contains up to  
$\mean{\hat{N}}\approx 5.4 \times 10^{7}$  photons.

\section{CONCLUSION}
We have shown that a two-mode single-atom laser 
can serve as a source of macroscopic entangled light.  
We identified two parameter regimes for 
which the quantum state of  the cavity field 
is entangled for a long period of time. 
For these parameters,  the master equation for the density operator 
of the two cavity modes can be approximately 
reduced  to the master equation 
for a nondegenerate parametric oscillator in the 
parametric approximation. 

The mean number of photons in the cavity field can 
be strongly increased if parameters beyond the 
parametric limit are chosen. This enhancement 
of the mean photon numbers is accompanied 
by a shortening of the time slice for which 
the entanglement criterion is fulfilled. 
As the initial state of the cavity field, we chose 
either  the vacuum  or  a coherent state. 
We demonstrated that the mean number of photons 
of the entangled cavity field can  increase by several orders of magnitude 
if  a coherent state instead of the vacuum is chosen as an initial state.

%\clearpage
\onecolumngrid
\appendix
\section{COEFFICIENTS \label{coefficients}}
Here we give the explicit definitions of the coefficients $\alpha_{ij}$ and $\beta_{ij}$ 
which enter the master equation~(\ref{master_field}) for the density operator $\rf$ 
of the  two cavity modes  
\begin{align}
%%%%%%%%%%%%%%%%%%%%%
\alpha_{11} = & 2 g_1^2 \Gamma _2 \left|\Omega _3\right|^2 \left|\Omega _4\right|^2 \left[ 4 \left(P_2^*+4 i
  \Delta_b \right) \left|\Omega _4\right|^2+P_1^* \left(4 \left|\Omega _3\right|^2+P_1
   \left(P_1+P_2^*\right)\right)\right]/(P_3 P_4) \label{a11}     \,, \\[0.2cm]   
   %%%%%%%%%%%%%%%%%%%%%%%%%%%%%%%%%%%%5
\beta_{11} = & -2 g_1^2 \Gamma _4 \left|\Omega _3\right|^2 \left|\Omega _4\right|^2  \left[ 4 P_1 \left|\Omega
   _4\right|^2+P_2^* \left(4 \left|\Omega _3\right|^2+P_1
   \left(P_1+P_2^*\right)\right)\right]/(P_3 P_4) \,, \\[0.2cm]
   %%%%%%%%%%%%%%%%%%%%%%%%%%
\alpha_{12} = &  -2 g_1 g_2 \Gamma _2 \Omega _3 \Omega _4\left|\Omega _3\right|^2 \left[ 4 P_1 \left|\Omega _3\right|^2+P_1^2
   \left(P_1+P_2^*\right)-4 \left|\Omega _4\right|^2 \left(2 P_1+P_2^*\right)\right]/(P_3 P_4) 
    \,,  \\[0.2cm]
    %%%%%%%%%%%%%%%%%%%%%%%%%%%%%%%%
\beta_{12} = & -2 g_1 g_2 \Gamma _4  \Omega _3 \Omega _4 \left|\Omega _4\right|^2 \left[ \left(P_1+P_2^*\right)
   \left|P_2\right|^2+4 \left|\Omega _4\right|^2 P_2+4 \left|\Omega _3\right|^2 \left(P_1-4 i
  \Delta_a\right)\right]/(P_3 P_4)      \,, \\[0.2cm]
   %%%%%%%%%%%%%%%%%%%%%%%%%%%%%%%5
\alpha_{22} = & 2 g_2^2 \Gamma _4  \left|\Omega _3\right|^2 \left|\Omega _4\right|^2 \left[ 4 \left(P_1^*+4 i
  \Delta_a\right) \left|\Omega _3\right|^2+P_2^* \left(4 \left|\Omega _4\right|^2+P_2
   \left(P_2+P_1^*\right)\right) \right]/(P_3 P_5)        \,,  \\[0.2cm]
   %%%%%%%%%%%%%%%%%%%%%%%%%%%%%%%%%%%%%%%%%%
\beta_{22} = & -2 g_2^2  \Gamma _2    \left|\Omega _3\right|^2 \left|\Omega _4\right|^2 \left[ 4 P_2 \left|\Omega
   _3\right|^2+P_1^* \left(4 \left|\Omega _4\right|^2+P_2
   \left(P_2+P_1^*\right)\right)\right] /(P_3 P_5)    \,,   \\[0.2cm]
   %%%%%%%%%%%%%%%%%%%%%%%%%%%%%%%%%%%%%%
\alpha_{21} = &-2 g_1 g_2  \Gamma _4 \Omega _3 \Omega _4 \left|\Omega _4\right|^2  
\left[ 4 P_2 \left|\Omega _4\right|^2 +P_2^2 \left(P_2+P_1^*\right)
-4 \left|\Omega_3\right|^2 \left(2 P_2+P_1^*\right) \right] 
/(P_3 P_5)      \,,  \\[0.2cm]
%%%%%%%%%%%%%%%%%%%%%%%%%%%%%
\beta_{21} = & -2 g_1 g_2 \Gamma _2\Omega _3 \Omega _4\left|\Omega _3\right|^2  \left[ \left(P_2+P_1^*\right)
   \left|P_1\right|^2+4 \left|\Omega _3\right|^2 P_1+4 \left|\Omega _4\right|^2 \left(P_2-4 i
  \Delta_b\right)\right] /(P_3 P_5) \label{b21}    \,.
   %%%%%%%%%%%%%%%%%%%%%%%%%%%%%%%%
\end{align}
The parameters $P_1$, $P_2$, $P_3$, $P_4$ and $P_5$ in Eqs.~(\ref{a11})-(\ref{b21}) are defined as 
\begin{align}
%%%%%%%%%%%%%%%%%%%%%%%%%%%%%%%%%%%
P_1 = &  \Gamma _3+\Gamma _4+2 i \Delta _b   \,, \\[0.2cm]
   %%%%%%%%%%%%%%%%%%%%%%%%%%%%
P_2 = & \Gamma _1+\Gamma _2+2 i \Delta _a \,, \\[0.2cm]
%%%%%%%%%%%%%%%%%%%%%%%%%%%%%%%%%%%%
P_3 = & \Gamma _2 \left|P_1\right|^2 \left|\Omega _3\right|^2
   +\Gamma _4    \left|P_2\right|^2 \left|\Omega _4\right|^2  
+8 \left|\Omega _3\right|^2 \left|\Omega _4\right|^2   \left(\Gamma _2+\Gamma _4\right) \,, \\[0.2cm]
%%%%%%%%%%%%%%%%%%%%%%%%%%%%%%%%%
P_4 = & 4 \left( \left|\Omega _3\right|^2 - \left|\Omega _4\right|^2\right)^2 
   +P_1 \left(P_1+P_2^*\right) \left|\Omega _3\right|^2
   +  P_2^* \left(P_1+P_2^*\right)\left|\Omega _4\right|^2 \,, \\[0.2cm]
%%%%%%%%%%%%%%%%%%%%%%%%%%%%%%%%%%%%%%%%
P_5 = & 4 \left( \left|\Omega _3\right|^2 - \left|\Omega _4\right|^2\right)^2 
   +P_1^* \left(P_2+P_1^*\right) \left|\Omega _3\right|^2
   +  P_2 \left(P_2+P_1^*\right)\left|\Omega _4\right|^2 \,.
%%%%%%%%%%%%%%%%%%%%%%%%%%%%%%%%%%%%%%
\end{align}

\section{CALCULATION OF THE MEAN VALUES \label{meanvalues}}
In the following, we outline the calculation of  the mean values that enter 
the total variance of the operators $\hat{u}$ and $\hat{v}$ in 
Eq.~(\ref{lhs_criterion}).  We begin with the mean values 
of  the quadrature operators 
defined in Eq.~(\ref{quadratures})  with respect to the 
density operator  $\rf$ of the two cavity modes. With the help of  Eq.~(\ref{master_field}), 
we derive the following system of differential equations for the mean values 
$\mean{b_1}$ and $\mean{b_2^{\dagger}}$, 
\begin{align}
\partial_t \left( 
\begin{array}{c}
\mean{b_1}\\[0.3cm]
\mean{b_2^{\dagger}}        
\end{array}
\right) = -
\left(
\begin{array}{ll}
C_{11} + \kappa_1 & C_{12} \\[0.3cm]
C_{21}^* & C_{22}^* + \kappa_2
\end{array}
\right)
\left( 
\begin{array}{c}
\mean{b_1}\\[0.3cm]
\mean{b_2^{\dagger}}        
\end{array}
\right)\,,
\end{align}
and 
$C_{ij} =  \alpha_{ij}+\beta_{ij}$. 
The solution to this set of coupled equations is
\begin{align}
\mean{b_1} = & e^{w_2 t}\left[ \cosh( w_1 t) \meanI{b_1} + 
\frac{1}{2 w_1} \left(\meanI{b_1} \left( C_{22}^* - C_{11}-\kappa_1+\kappa_2\right)
-2 \meanI{b_2^{\dagger}} C_{12}\right) \sinh(w_1 t)\right] \\
\mean{b_2^{\dagger}} = & e^{w_2 t}\left[ \cosh( w_1 t) \meanI{b_2^{\dagger}} + 
\frac{1}{2 w_1} \left(\meanI{b_2^{\dagger}} \left( C_{11} - C_{22}^*  +\kappa_1-\kappa_2\right)
-2 \meanI{b_1} C_{21}^* \right) \sinh(w_1 t)\right] \,,
\end{align}
where 
\be
w_1 =  \frac{1}{2}\sqrt{4 C_{12} C_{21}^* +(C_{11}-C_{22}^* +\kappa_1-\kappa_2)^2}  \, , \qquad
w_2 =  -\frac{1}{2} (C_{11} + C_{22}^* +\kappa_1 +\kappa_2 ) \, ,
\ee
and   $\meanI{\,\cdot\,} = \mean{\,\cdot\,}(t=0)$  denotes  the initial mean value  at $t=0$. 
Note that the mean values $\mean{b_1^{\dagger}}$ and  $\mean{b_2}$ can 
be obtained from $\mean{b_1}$  and $\mean{b_2^{\dagger}}$ by complex conjugation, 
i.e. $\mean{b_1^{\dagger}} = \mean{b_1}^*$ and  $\mean{b_2} = \mean{b_2^{\dagger}}^*$. 
%%%%%%%%%%%%%%%%%%%%%%%%%

The remaining mean values in Eq.~(\ref{lhs_criterion}) involve products of the 
operators $b_i$ and $b_i^{\dagger}$. 
With the aid of  Eq.~(\ref{master_field}), we  obtain the following set of differential equations, 
\be
\partial_t \mf{R} = M \mf{R} + \mf{I} \, , 
\label{meanval2}
\ee
where 
$\mf{R} = \left(
\mean{b_1^{\dagger} b_1}  ,\,
\mean{b_2^{\dagger} b_2}  ,\,
\mean{b_1 b_2} ,\,
\mean{b_1^{\dagger} b_2^{\dagger}} 
\right)\,$ 
and
\begin{align}
M = - \left(
\begin{array}{l@{\hspace*{0.5cm}}l@{\hspace*{0.5cm}}l@{\hspace*{0.5cm}}l}
 D_{11} & 0 & C_{12}^* &  C_{12} \\[0.3cm]
0 & D_{22}  &  C_{21}^* &  C_{21} \\[0.3cm]
C_{21} & C_{12} &  D_{12} & 0 \\[0.3cm]
C_{21}^*  & C_{12}^* & 0 &  D_{12}^* 
\end{array}
\right)
\quad,\quad
\mf{I} = - \left(
\begin{array}{c}
\beta_{11} + \beta_{11}^* \\[0.3cm]
\beta_{22} + \beta_{22}^* \\[0.3cm]
\alpha_{12} + \alpha_{21} \\[0.3cm]
\alpha_{12}^* + \alpha_{21}^*
\end{array} 
\right) \,.
\label{m_and_i}
\end{align}
The elements of the matrix $M$  are defined as 
\be
C_{ij} =  \alpha_{ij}+\beta_{ij} \,,\quad
D_{ii} =  \alpha_{ii} +\alpha_{ii}^*+\beta_{ii} +\beta_{ii}^*  +2\kappa_i \,, \text{ and } 
D_{12} =  C_{11} + C_{22} +\kappa_1 +\kappa_2 \,.
\ee
The differential equation Eq.~(\ref{meanval2}) can be solved numerically without 
difficulties. An analytical solution can be obtained, for example, by means of the Laplace transform method 
which yields the following results for the components $R_i$ of the vector $\mf{R}$, 
\be
R_i = \sum\limits_{k=1}^{4} \big[ \text{Res} ( f_i , \lambda_k  ) \,+\,
\text{Res} ( g_i , \lambda_k  ) \big]e^{\lambda_k t}
 + \text{Res} ( g_i, 0 ) \,.
\ee
In this equation, expressions of the type $\text{Res}(h, z )$ denote the residue of the 
function $h$ evaluated at $z$, and the functions $\mf{f} = \left(f_1 , \, f_2, \, f_3, \, f_4\right)\,$  
and $\mf{g} = \left( g_1  ,\,g_2   ,\,g_3  ,\,g_4  \right)$  are determined by 
\be
\mf{f}(s) = \left[ s\ \hat{1}_4 -M\right]^{-1} \mf{R}_0 \qquad \text{and} \qquad \mf{g}(s) = \left[ s\ \hat{1}_4 -M\right]^{-1} (\mf{I}/s)\,,
\ee
respectively. Here $\hat{1}_4$ denotes the $4\times4$ identity matrix, and the vector $\mf{R}_0$ is 
the initial value of $\mf{R}$ at $t=0$,  
$\mf{R}_0 = \left(
\mean{b_1^{\dagger} b_1}_0  ,\,
\mean{b_2^{\dagger} b_2}_0  ,\,
\mean{b_1 b_2}_0 ,\,
\mean{b_1^{\dagger} b_2^{\dagger}}_0  
\right)$. 
Finally, the parameters $\lambda_k$   are the four (complex) eigenvalues of 
the matrix $M$ which is defined in Eq.~(\ref{m_and_i}). The eigenvalues $\lambda_k$  can 
be obtained as  the roots of the following equation, 
\begin{align}
 & s^4 + \left[ D_{11} +  D_{22} +  D_{12} +D_{12}^*   \right] s^3
+ \left[ |D_{12}|^2-2 C_{21} C_{12}^{*}-2 C_{12} C_{21}^{*}+D_{11} D_{22}+ 
   (D_{11}+D_{22}) (D_{12}+D_{12}^*)  \right] s^2 \notag \\[0.2cm]
& + \left[(D_{11}+D_{22}) |D_{12}|^2  -(C_{21} C_{12}^{*} + C_{12} C_{21}^{*})
   (D_{11}+D_{22}+D_{12} +D_{12}^* ) + D_{11} D_{22}(D_{12} +D_{12}^*)  \right] s  \notag \\[0.2cm]
& + C_{21}^2 (C_{12}^{*})^2 - 
(2 C_{12} C_{21}^{*}+D_{11} D_{12}+D_{22} D_{12}^{*}) C_{21} C_{12}^{*}
+(C_{12} C_{21}^{*}-D_{22} D_{12}) (C_{12}
   C_{21}^{*}-D_{11} D_{12}^{*})  \,\, = \,\,0 \,.
\end{align}

\twocolumngrid

\end{document}